\documentclass[%
 reprint,
superscriptaddress,
 amsmath,amssymb,
 aps,
]{revtex4-2}

\usepackage{graphicx}
\usepackage{xcolor}
\usepackage{dcolumn}
\usepackage{bm}
\usepackage{braket}
\usepackage[none]{hyphenat}
\usepackage{comment}
\usepackage[colorlinks=true,linkcolor=darkblue,urlcolor=darkblue,filecolor=darkblue,citecolor=darkblue,linktocpage=true
      ]{hyperref}
\definecolor{darkblue}{rgb}{0.2, 0, 0.8}

\renewcommand{\[}{\begin{equation}\begin{aligned}}
\renewcommand{\]}{\end{aligned}\end{equation}}

\newcommand{\bea}{\begin{eqnarray}}
\newcommand{\eea}{\end{eqnarray}}

\begin{document}

\title{Thermodynamics of Kerr Effective Black Hole Geometries }

\author{Nelson Hernández Rodríguez}
\email{nelsonhdez21@gmail.com}
\affiliation{%
 Instituto de Fisica Teorica UAM/CSIC, 28049 Madrid, Spain.\\
}

\author{Maria J. Rodriguez}%
 \email{maria.rodriguez@gmail.com}

 \affiliation{
Department of Physics, Utah State University, Logan, UT 84322, USA.\\
}
 \affiliation{
Black Hole Initiative, Harvard University, Cambridge MA 02138, USA.\\
}
\affiliation{%
 Instituto de Fisica Teorica UAM/CSIC, 28049 Madrid, Spain.\\
}

\date{\today}

\begin{abstract}

We uncover the thermodynamical properties of a class of non-asymptotically flat geometries, referred here as the Kerr effective geometries, that realize the hidden symmetries of Kerr black hole
spacetimes via Teukolsky’s equation 
in the theory of linear gravitational perturbations. While the thermodynamic properties, such as entropy, remain the same as those of Kerr black holes, the asymptotic charges, as we will demonstrate, are different. In this new framework with the ambient flat  asymptotic boundary removed, we derive the mass and angular momentum of the Kerr effective geometries using the regulated Komar integral and the Brown-York prescription. We also obtain explicit expressions for the corresponding gauge potentials and electric charges. By deriving the asymptotic charges, we demonstrate that both the first law of black hole mechanics and the Smarr law hold.

\end{abstract}

\maketitle


\section{Introduction}

The internal dynamics of analytical models for the gravitational tidal deformation of black holes are often complex and intricate. Love numbers quantify the response of a black hole to external gravitational forces. They are critical in studying binary black hole systems, especially during inspirals and mergers.
Recent advancements in the study of tidal deformations have revealed that static Love numbers vanish to linear \cite{jakobsen2023, damour2009, binnington2009, fang2005, kol2012, Chakrabarti:2013lua, gurlebeck2015, 
porto2016, letiec2021a, letiec2021b, chia2021, charalambous2021} and potentially even more surprising at higher nonlinear orders \cite{Riva:2023rcm}. The vanishing of Love numbers in the context of static perturbations at linear order is explained by its association with the $SL(2, R)$ representation theory. According to the proposed explanation  \cite{Charalambous:2021mea, Hui:2021vcv} (see also  \cite{Chia:2020yla,Charalambous:2021kcz,Charalambous:2022rre,BenAchour:2022uqo,Hui:2022vbh}), it is the highest weight property that forces the static Love numbers to vanish. These models rely in certain effective Kerr geometries (KEG)  that realize the hidden symmetries of nature captured by the mass-less scalar perturbations around a Kerr black hole. Specifically, one proposal \cite{Hui:2021vcv} is characterized by an $SO(4,2)$ hidden symmetry, represented by the Starobinsky KEG, while the other \cite{Charalambous:2021mea} is characterized by an $SL(2,R) \times U(1)$ symmetry, associated with the so-called Love KEG.
The lack of uniqueness among these two proposals poses a significant issue as each one is associated with distinct hidden symmetries that
are likely unrelated to one another. The observation of the present paper is to quantify the differences among the effective geometries by contrasting their physical asymptotic thermodynamic quantities and mechanical laws.

Previous studies, which have focused on explaining the rigidity of Kerr black holes using hidden symmetries, have predominantly relied on non-dynamical models. An important aspect of the physical processes in black hole mergers in General Relativity is the motion of the extended bodies.  Recent advancements have addressed this issue \cite{Perry:2023wmm,Charalambous:2022rre,Saketh:2023bul}. In dynamic scenarios, an $SL(2,R)\times SL(2,R)$ hidden symmetry seems to realize the non-zero gravitational Love numbers at low frequencies $\omega \ll 1$ \cite{Perry:2023wmm}. The associated effective geometry, known as subtracted geometries, was first identified in \cite{cvetic:2011dn} and is based on the realization of $SL(2,R)\times SL(2,R)$ generators \cite{Castro:2010fd}. It was later revealed in \cite{Cvetic:2012tr} that the previously obtained subtracted geometry of four-dimensional rotating black holes, whose massless wave equation exhibit $SL(2,R) \times SL(2,R)$ symmetry, can be derived from certain asymptotically flat multi-charged rotating black holes through a suitable scaling limit. In this process, the asymptotic boundary of general asymptotically flat black holes can be modified such that an asymptotically conformal symmetry emerges. These modified black holes, with their asymptotic geometry removed in this manner, satisfy the equations of motion of the so-called STU model -- a four-dimensional N=2 supergravity coupled to three vector fields. When extended to five spacetime dimensions, this configuration corresponds to minimal supergravity.

Notably, the Love and Starobinsky KEGs, relevant in the context of tidal Love numbers, are also solutions of the STU model, as recently discovered in \cite{Cvetic:2024dvn}. The scaling construction employed in this proof preserves all near-horizon properties of the black holes, including the thermodynamic potentials and entropy.

Asymptotically, all KEG spacetime metrics exhibit a distinct structure compared to flat spacetime. The boundary geometry can be interpreted as black holes confined asymptotically within a box \cite{Cvetic:2012tr, Cvetic:2024dvn}, resembling the behavior observed in asymptotically AdS spacetimes. One may wonder what are the specific asymptotic charges such as mass and angular momentum in the KEG? Are these the same as those in Kerr black holes, in the same way that physical properties defined on the horizon are preserved?

The thermodynamics of the KEGs has not been extensively investigated; progress in this area could serve as a foundation for exploring gravity/field theory duality in asymptotically boxed spacetimes.

In this letter we focus on the study of global thermodynamic properties of the Love and Starobinsky KEG (respectively exhibiting $SL(2, R) \times U (1)$ and $SO(4, 2)$ hidden symmetries).
We demonstrate that certain subtleties arise in deriving the mass and asymptotic charges. Furthermore, we clarify the origin and geometric significance of the KEG geometries within the framework of the black hole mechanics laws governing these metrics.

The KEG are realized as solutions of N=2 supergravity \cite{Cvetic:2024dvn}, and the corresponding Lagrangian density of this theory is given by 
\begin{align}
{\cal L}_4 =& R\, {*{\bf 1}} - \frac{1}{2} {*d\varphi_i}\wedge d\varphi_i - \frac{1}{2} e^{2\varphi_i}\, {*d\chi_i}\wedge d\chi_i\\
& - \frac{1}{2} e^{-\varphi_1}\,( e^{\varphi_2-\varphi_3}\, {*  F_{1}}\wedge   F_{ 1}\nonumber + e^{\varphi_2+\varphi_3}\, {*   F_{ 2}}\wedge   F_{ 2}\nonumber\\
  &+ e^{-\varphi_2 + \varphi_3}\, {*  {\cal F}_1 }\wedge   {\cal F}_1 + e^{-\varphi_2 -\varphi_3}\, {* {\cal F}_2}\wedge   {\cal F}_2)\nonumber\\
&- \nonumber \chi_1\, (  F_{1}\wedge  {\cal F}_1 + 
                   F_{ 2}\wedge  {\cal F}_2)\,,
\label{d4lag}
\end{align}
where the index $i$ appearing in the dilatons, $\varphi_{i}$, and axions, $\chi_{i}$, takes values from 1 to 3. The dilaton and axion scalar fields are respectively denoted as $\varphi_{i}$ and $\chi_{i}$. 
The $U(1)$ field strengths can be expressed in terms of potentials as follows 
\bea
  F_{(2)1} &=& d   A_{(1)1} - \chi_2\, d {\cal A}^2_{(1)} , \qquad   {\cal F}^2_{(2)} = d  {\cal A}^2_{(1)} \nonumber\\
  F_{(2) 2} &=& d  A_{(1) 2} + \chi_2\, d {\cal A}^1_{(1)} - \chi_3\, d   A_{(1) 1} +\chi_2\, \chi_3\, d  {\cal A}^2_{(1)} , \nonumber\\
  {\cal F}^1_{(2)} &=& d  {\cal A}^1_{(1)} + \chi_3\, d  {\cal A}^2_{(1)} , \nonumber.
\eea
 The rotating KEG metric is of the form 
\bea\label{ACmetric}
ds^{2}&=&-\Delta_{0}^{-1/2} G_{0} \, (dt+A_{0}\,d\phi)^{2} \\
&&+ \Delta_{0}^{1/2}\left(\frac{dr^{2}}{\Delta}+d\theta^{2} + \frac{\Delta}{G_{0}}\sin^{2}\theta\, d\phi^{2}\right) \nonumber
\eea
with arbitrary $\Delta_0$, the black hole horizons $r=r_\pm$ defined by $\Delta(r_\pm)=0$ ($r_+>r_->0$) and,
\bea
\left. A_0\,\, \right|_{r=r_+}=- (r_+^2+a^2)\,,\,\, \left. G_0\,\, \right|_{r=r_+}= -a^2\sin^2\theta\,,
\eea
such that effectively the near black hole horizon Kerr region is preserved and the corresponding asymptotic form of the geometry reads
\begin{equation}\label{ACmetric1}
ds_{AC}^{2}=-Y^{2p} dt^2+B^{2}\frac{dY^2}{Y^{2s}}+Y^{2q}(d\theta^2+\sin^2\theta d\phi^2)\,,
\end{equation}
where $(p,q,s,B)$ are constants; in particular KEGs are not asymptotically flat implying $p,s\ne0$, and $q,B\ne1$.\\

{\bf Starobinsky KEG} One of the metrics that has gained increasing interest for its fundamental role in describing Love symmetries is the one found 
\cite{Hui:2022vbh}, has a ``Starobinsky" hidden symmetry $SO(4,2)$. Similar to the previous cases, the line element will be \eqref{ACmetric}; however, the functions will depend on how we perform our effective near-zone approximation. Specifically, we consider
\begin{align}\label{equs}
G_{0} & = \Delta -a^{2}\sin^{2}\theta, \qquad
\Delta =r^{2}+a^{2}-2mr , \nonumber \\
\Delta_{0} & = (r_{+}^{2}+a^{2})^{2},\qquad
A_{0} = \frac{a \sin^{2}\theta}{G_{0}}(r_{+}^{2}+a^{2}).
\end{align}
The metric is a conformally flat spacetime, has vanishing Ricci scalar (though not Ricci tensor), and represents a solution supported by external fields as we describe next. Starting from the general STU solution \cite{Chong:2004na,Cvetic:2012tr} parameterized by the mass parameter $\bar{m}$, the rotational parameter $\bar{a}$ and four charge parameters $\bar{\delta}_I$ ($I = 1, 2, 3, 4$) a scaling limit can be employed to define the fields supporting the KEG.
 This procedure was employed by Cvetic et al. \cite{Cvetic:2012tr} in the context of the geometry exhibiting $SL(2,R)\times SL(2,R)$ symmetry of the wave equation \cite{Castro:2012av}, known as subtracted geometry. 
Considering the four large charge solutions \cite{Chong:2004na,Cvetic:2012tr} with $\bar{\delta}_1 = \bar{\delta}_2 = \bar{\delta}_3 \equiv \bar{\delta}$ and $\bar{\delta}_4 \equiv \bar{\delta}_0$ but denoting all the quantities taken with “barred” notation we take
\bea
\label{scaling1}
\nonumber\bar{r}= r \epsilon \,,\qquad \bar{t} =t \epsilon^{-1}\,,\qquad  \bar{\phi}= \epsilon \phi \,,\qquad \bar{m}=m \,\epsilon\,,\\
 \bar{a}=a \,\epsilon\,, \qquad
 \sinh\bar{\delta} = \sinh \bar{\delta_0} \equiv   \epsilon^{-1/2} \sinh{\delta} 
\eea
 with scaling parameter $\epsilon\rightarrow 0$, provided one identifies $\sinh^4{\delta} \equiv r_+/(r_++r_-)$. In terms of the new unbarred coordinates, the metric \eqref{ACmetric} becomes that of Starobinsky with the functions defined in \eqref{equs}. The sources supporting this geometry can also be obtained in this scaling limit, resulting in
\bea
\chi_1=\chi_2=\chi_3=0,\qquad {\varphi_1}={\varphi_2}={\varphi_3}= 0\,,
\eea

\begin{align}\label{gaugefield1}
   A_t\, dt \equiv A_{(1)2}^{Star}=(\mathcal{A}_{(1)}^{2})^{Star}= \frac{(r-m)}{2 m \sinh^{2}\delta}\,dt 
\end{align}
and
\begin{align}\label{gaugefield2}
    A_{(1)1}^{Star}= \mathcal{A}_{(1)1}^{Star}=\frac{a\, \cos\theta }{2 m \sinh ^2\delta} \, dt -2 m \cos\theta \sinh^2\delta \,d\phi \; .
\end{align}

The asymptotic charges can easily be obtained from the Gauss law
\bea
\label{eq:ChargeStar}
Q=\mathcal{Q} =\frac{1}{2} m \,\sinh^2\delta\,.
\eea

{\bf Love KEG} 
 We will also work with the KEG presented in \cite{Charalambous:2021kcz}. In this particular case, the hidden "Love" symmetry possessed by this metric corresponds to $SL(2,R)\times U(1)$. 
Once again, the line element is defined by \eqref{ACmetric}, albeit in this case the functions are
\begin{align}\label{equs1}
G_{0} & = \Delta -a^{2}\sin^{2}\theta, \,\,\, \Delta =r^{2}+a^{2}-2mr,\nonumber \\
A_{0} &= \frac{a \sin^{2}\theta}{G_{0}}(r_{+}^{2}+a^{2}+\beta(r-r_{+})), \nonumber \\
\Delta_{0} & = (r_{+}^{2}+a^{2})^{2}(1+\beta^{2}\Omega^{2}\sin^{2}\theta),
\end{align}
where $\Omega$ is the black hole angular velocity, and the parameter $\beta = \frac{1}{2\pi T}$, related to the black hole Bekenstein-Hawking temperature $T$ (see definitions below). The corresponding Ricci scalar exhibits a divergence at $\Delta_0=0$, indicating a curvature singularity in the metric.
A scaling limit with slightly different formulas from those used previously can be applied to the general STU solution \cite{Chong:2004na,Cvetic:2012tr}. The limit is implemented by means of the following scaling
\bea
\label{scaling2}
 &\bar{r}= r \epsilon \,,\qquad \bar{t} =t \epsilon^{-1}\,,\qquad  \bar{m}=m \,\epsilon\,,\\
& \bar{a}=a \,\epsilon\,, \qquad
 \sinh\bar{\delta} = -\sinh \bar{\delta_0} \equiv  \epsilon^{-1/2} \sinh{\delta} . \nonumber
\eea
 We recover the Love geometry in the limit upon setting $\sinh^4{\delta}=r_+/(r_+-r_-)$.
This scaling limit helps in explicitly identifying the sources enabling the Love KEG as a solution of $N=2$ supergravity. The corresponding sources are given by:
\bea
&&\chi_1=-\chi_2=\chi_3=-\frac{a}{m} \cos\theta\,,\\
&& e^{\varphi_1}=e^{\varphi_2}=e^{\varphi_3}=\frac{m}{\sqrt{m^2-a^2\cos^2\theta}}\,,
\eea
and the gauge potentials by:
\begin{align}\label{gauge1}
    A_{(1)2}^{Love}=-(\mathcal{A}_{(1)}^{2})^{Love}&& = \frac{m(m-r)}{2 \sinh^{2}\delta\, (m^2-a^{2}\cos^2\theta)} \, dt \nonumber\\
    &&-\frac{2 a m^2  \,\sinh ^2\delta \sin^2\theta}{(m^2-a^2 \cos ^2\theta)} \, d\phi
\end{align}
and
 \begin{align}\label{gauge2}
    A_{(1)1}^{Love}= (\mathcal{A}_{(1)}^{1})^{Love}&& =-\frac{a(m-r) \cos\theta\hspace{0.1cm}}{2 \sinh^{2}\delta\, (m^2-a^{2}\cos^2\theta)} \, dt \nonumber\\
    &&-\frac{2 m (m^2-a^2)  \,\sinh ^2\delta \cos\theta}{(m^2-a^2 \cos ^2\theta)} \, d\phi .
\end{align}
We can now write the electric charges
\bea
\label{eq:ChargeLove}
Q=-\mathcal{Q} =\frac{1}{2} m \,\sinh^2\delta\,.
\eea
An important point to notice here is that the dilatons and axions have a $\theta$-dependence.
Interestingly, as for the subtracted KEG in \cite{Cvetic:2012tr}, the gauge fields supporting such KEG geometries vary as a function of radial distance at infinity. This fact will become relevant in the definitions of the asymptotic charges.

\section*{Thermodynamics}

Black holes behave as thermodynamic objects, with their thermodynamic properties determined by the behavior of their geometry at certain specific boundaries. All KEG possess the same thermodynamic quantities defined on the black hole event horizon $r=r_+$, including the entropy, temperature and angular velocity
\bea
S= \pi (r_{+}^{2}+a^{2})\,,\,\,\, T\equiv \frac{r_+-r_-}{8\pi M r_+}\,,\,\,\, \Omega=\frac{a}{2M r_+}\,,
\eea
which also remain consistent with those of Kerr. 
However, the mass and angular momentum intricately rely on the asymptotic properties of the asymptotic geometry. Let us start by analyzing the mass of the KEG geometry. In this context, we specifically consider the Starobinsky KEG \eqref{ACmetric} with \eqref{equs}. A similar procedure can be adapted for the Love KEG. 
The geometry on a surface with $r = const.$ can locally be cast in ADM-like form 
\bea
&&h_{ij} dx^i dx^j \nonumber\\
&&= -N^2 dt^2 + \sigma_{ab} (dy^a + N^a dt)(dy^b + N^b dt),
\eea
where \(N\) and \(N^a\) are the lapse function and the shift vector respectively, and the \(y^a\) are the intrinsic coordinates on the closed surfaces \(\Sigma\).
Through a rigorous examination of the quasilocal formalism (see \cite{Astefanesei:2009wi} for a review), notably advanced by Brown and York , we extract the relevant information about the properties of the conserved charges of KEG black holes. Physically, this method poses on the idea that a collection of observers on the hypersurface whose boundary metric is $h_{ij}$ all observe the same value of the conserved charge 
\begin{equation}
Q_\xi=\int_{\Sigma} d\theta d\phi \hspace{0.1cm} \sqrt{\sigma}\left( n^{i}\tau_{ij}\xi^{j}\right) ,
\end{equation}
provided this surface has an isometry generated by $\xi^i$. Here $n^{i}$ is a normal vector to a surface of $r=const.$, $\sqrt{\sigma}$ is the determinant of the 2D-boundary metric. We use the divergence-
free boundary stress tensor proposed in employed in e.g. \cite{Astefanesei:2009mc}
\bea
\tau_{ij}  = \frac{1}{8\pi G} \left( K_{ij} - h_{ij} K - \Psi(R_{ij} - R h_{ij})
\right. \nonumber &\\
 \left. - h_{ij} \nabla^2 \Psi + \nabla_i \nabla_j \Psi \right) & 
\eea
where $\Psi = c/\sqrt{R}$, $K$ is the extrinsic curvature , $h_{ij}$ is the induced boundary metric, and $\mathcal{R}_{ij}$ its Ricci tensor. The constant $c$ that enters the above relation depends on the boundary topology — for Kerr one finds $c=\sqrt{2}$ and $c=1/2$ for the KEG. Calculating the Brown-York mass, we get
\bea
M_{BY}=m\,.
\eea

Subsequently, our objective is to assess the coherence of the results obtained through the Brown-York formalism with the Komar mass formula. In this formalism, the mass is defined through the following Komar integral:
\begin{equation}
    M_K=\frac{1}{8\pi}\int_{S}\star \textrm{d}\xi,
\end{equation}
related to the timelike Killing vector $\xi=\partial t$, where $\star$ is the Hodge star operator and $S$ is the 2-surface boundary. One can show that
\begin{equation}
    M_K=r-m+\mathcal{O}\left(\frac{1}{r^2}\right),
\end{equation}
that diverges linearly with r. We can, however, show that this divergence gets regulated once we take the asymptotic gauge fields and charges into account. Employing the same regularization method as employed in \cite{Cvetic:2014nta}, one can define \(H^{\mu\nu} = \nabla^\mu \zeta^\nu (t) - \nabla^\nu \zeta^\mu (t)\) which allows us to show that:
\[
\nabla_\mu H^{\mu\nu} = -16\pi(T^\nu_\mu - \frac{1}{2}T\delta^\nu_\mu)\zeta^\mu (t) ,
\]
the mass has in addition an electromagnetic contribution $T^{\mu\nu}$.
For the case of static KEG, we obtain:
\[
\nabla_r (H^{rt} + 8\pi[4 F^{rt}A_t(r)]) = 0 . 
\]
Note that as \(r \to \infty\), the only contribution comes from the term with \(A_t(r)\) that was previously identified \eqref{gaugefield1}. We obtain:
\[
M_{K\text{reg}} = M_K + 4 Q A_t(r) = M_K + (2 m - r), 
\]
where \(M_K\) is the unregulated Komar mass and the charge parameter defined by \(r^2F^{rt} = Q\). The terms linear in \(r\) cancel, and the regulated Komar mass an then be easily obtained
\[\label{finmass}
M \equiv M_{K\text{reg}} = M_{BY} . 
\]
In other words, the Komar formula yields identical results to those obtained using the Brown-York formalism.

We can move to define the angular momentum. The Komar formalism provides an equivalent approach related to the rotational Killing vector $\eta=\partial\phi$. The angular momentum is defined through the following Komar integral, 
\begin{equation}
    J_K=\frac{1}{16\pi}\int_{S}\star \textrm{d}\eta,
\end{equation}
which differs by a factor of 2 from the previously obtained integral for the mass.
Employing these methods we find that angular momentum is defined as
\bea\label{finmom}
J\equiv J_K=J_{BY}=0.
\eea
To calculate the gauge potentials, we employ equation (6.6) from \cite{chow2014black}, adjusting it accordingly for the scaling limit of the KEG.
Then, for the Starobinsky KEG, the gauge potential on the black hole horizon $r=r_+$ can be
written as
\bea\label{eq:potential1}
\Phi(r_+)=\frac{m-r_+}{2 m \sinh^2\delta}
\eea
 evaluated at $\theta=0$. 
Having defined all the thermodynamic quantities for the geometries, we can easily show that the Smarr law continues to hold, such that
\bea
M= 2 (TS + \Omega \, J)+  4 \Phi(r_+) Q\,,
\eea
as well as the first law
\bea
dM=  T\, dS + \Omega \,dJ + 4 \Phi(r_+)\,dQ \,.
\eea

 It is worth emphasizing that the gauge fixing procedure is central for this laws to be obeyed. In this case, the gauge was uniquely fixed by the scaling limit, as we will  now discuss.

Another approach to deriving these thermodynamic laws for the KEG involves applying the scaling limit to Smarr's Law and the first law of thermodynamics for the original solution \eqref{ACmetric} of $N=2$ SUGRA (for details see \cite{Chong:2004na}).

In the case of Starobinsky KEGs the scaling limit is implemented through the rescalings \eqref{scaling1} for the mass and angular momentum. All thermodynamic quantities defined on the horizon mirror those in the asymptotically flat black hole solutions. Matching the asymptotic charges is a more intricate process. The mass of the STU four-charge black hole is given by \(\tilde{M}=(\tilde{m}/4)\sum_i \cosh 2 \tilde{\delta}_i\). Utilizing the scaling limit \eqref{scaling1}, we find for the Starobinsky KEG that
\begin{equation}\label{massscaling}
\frac{M}{\epsilon} = m + \frac{2 m}{\epsilon} \sinh ^2\delta.
\end{equation}
This a priori seems to indicate a divergence. However, simultaneously, the gauge potentials acquire an infinite constant term omitted in \eqref{eq:potential1}. Along with the charges, the divergent pairs cancel the divergent term in \eqref{massscaling}. We conclude that with the rescaling method the mass and angular momentum for the Starobinsky KEG are respectively \eqref{finmass} and \eqref{finmom}. This verification confirms the accuracy of the previous Komar and BY definitions. The scaling procedure also uniquely determines the gauge of the fields \eqref{eq:potential1} and electric charges \eqref{eq:ChargeStar}.

To complete the investigation into the thermodynamics of Love KEG black holes, we employ the scaling limit \eqref{scaling2}. We are able to determine the mass and angular momentum, yielding:
\bea
M^{Love}= m \,,\qquad J^{Love}= a \, \beta / 2.
\eea
In addition to the first law, the Love KEG also fulfills the Smarr relation.
\bea
M^{Love}= 2 (TS + \Omega \, J^{Love})+ 3 \Phi(r_+) Q+  \tilde{\Phi}(r_+) \mathcal{Q}\,,
\eea
 with the relevant gauge potentials evaluated on the black hole horizon $r=r_+$ and $\theta=0$ given by 
 \bea\label{eq:potential2}
 \Phi(r_+)=-\tilde{\Phi}(r_+)\equiv \frac{m}{2 (m-r_+) \sinh^2\delta}
 \eea
 and electric charges specified in \eqref{eq:ChargeLove}.

\section*{Discussion and Conclusion}

We have explicitly identified the asymptotic charges for the KEGs. To our knowledge, their thermodynamics had not been explored in detail. It is important to emphasize that a consistent extension of the black hole thermodynamic equations to KEGs relies significantly on recognizing and properly considering their asymptotic falloffs. KEGs have a very different asymptotic structure compared to asymptotically flat and asymptotically AdS black holes. Considering their asymptotic configuration, we were able to demonstrate equivalent expressions for the mass and angular momentum derived from the regulated Komar Integral and the Brown-York prescription. 

KEGs do not a priori satisfy the black hole thermodynamic relations. By consolidating the definitions of the gauge fields in \cite{Cvetic:2024dvn}, we managed to verify the electric charges and gauge potentials necessary to establish the first law of black hole mechanics. Our analysis showed that the Smarr formula and the first law of black hole mechanics hold.

Remarkably, for KEGs, the thermodynamic quantities defined on the horizon appear insensitive to the environmental context in which they are embedded. The results for the entropy, temperature, and angular velocity of KEGs exhibit universal attributes similar to those observed in Kerr black holes. The universality seems to be related to these quantities being defined on the horizon, thereby reinforcing the proposition of the existence of microscopic states through the entropy of black holes. However, the asymptotic expressions for the charges and gauge potentials capture the "box" in which these systems are placed. As we demonstrated here, the asymptotic structure of KEGs is crucial in establishing the first law of black hole mechanics and Smarr law.\\

{\bf Acknowledgments} 
The authors would like to thank Mirjam Cvetic and Oscar Varela for the enlightening discussions.
We gratefully thank the Mitchell Family Foundation at Cook's Branch workshop and the Centro de Ciencias de Benasque Pedro Pascual for their warm hospitality.
NHR was supported by RYC-2016-21159 and CNS2022-135880. MJR was partially supported through the NSF grant PHY-2309270, CNS2022-135880, CEX2020-001007-S and PID2021-123017NB-I00, funded by MCIN/AEI/10.13039/501100011033 and by ERDF A way of making Europe.

\bibliography{bibliography}

\providecommand{\noopsort}[1]{}\providecommand{\singleletter}[1]{#1}%
\begin{thebibliography}{32}%
\makeatletter
\providecommand \@ifxundefined [1]{%
 \@ifx{#1\undefined}
}%
\providecommand \@ifnum [1]{%
 \ifnum #1\expandafter \@firstoftwo
 \else \expandafter \@secondoftwo
 \fi
}%
\providecommand \@ifx [1]{%
 \ifx #1\expandafter \@firstoftwo
 \else \expandafter \@secondoftwo
 \fi
}%
\providecommand \natexlab [1]{#1}%
\providecommand \enquote  [1]{``#1''}%
\providecommand \bibnamefont  [1]{#1}%
\providecommand \bibfnamefont [1]{#1}%
\providecommand \citenamefont [1]{#1}%
\providecommand \href@noop [0]{\@secondoftwo}%
\providecommand \href [0]{\begingroup \@sanitize@url \@href}%
\providecommand \@href[1]{\@@startlink{#1}\@@href}%
\providecommand \@@href[1]{\endgroup#1\@@endlink}%
\providecommand \@sanitize@url [0]{\catcode `\\12\catcode `\$12\catcode `\&12\catcode `\#12\catcode `\^12\catcode `\_12\catcode `\%12\relax}%
\providecommand \@@startlink[1]{}%
\providecommand \@@endlink[0]{}%
\providecommand \url  [0]{\begingroup\@sanitize@url \@url }%
\providecommand \@url [1]{\endgroup\@href {#1}{\urlprefix }}%
\providecommand \urlprefix  [0]{URL }%
\providecommand \Eprint [0]{\href }%
\providecommand \doibase [0]{http://dx.doi.org/}%
\providecommand \selectlanguage [0]{\@gobble}%
\providecommand \bibinfo  [0]{\@secondoftwo}%
\providecommand \bibfield  [0]{\@secondoftwo}%
\providecommand \translation [1]{[#1]}%
\providecommand \BibitemOpen [0]{}%
\providecommand \bibitemStop [0]{}%
\providecommand \bibitemNoStop [0]{.\EOS\space}%
\providecommand \EOS [0]{\spacefactor3000\relax}%
\providecommand \BibitemShut  [1]{\csname bibitem#1\endcsname}%
\let\auto@bib@innerbib\@empty
\bibitem [{\citenamefont {Jakobsen}\ \emph {et~al.}(2023)\citenamefont {Jakobsen}, \citenamefont {Mogull}, \citenamefont {Plefka},\ and\ \citenamefont {Sauer}}]{jakobsen2023}%
  \BibitemOpen
  \bibfield  {author} {\bibinfo {author} {\bibfnamefont {G.~U.}\ \bibnamefont {Jakobsen}}, \bibinfo {author} {\bibfnamefont {G.}~\bibnamefont {Mogull}}, \bibinfo {author} {\bibfnamefont {J.}~\bibnamefont {Plefka}}, \ and\ \bibinfo {author} {\bibfnamefont {B.}~\bibnamefont {Sauer}},\ }\href@noop {} {\bibfield  {journal} {\bibinfo  {journal} {Phys. Rev. D}\ }\textbf {\bibinfo {volume} {131}},\ \bibinfo {pages} {011603} (\bibinfo {year} {2023})},\ \Eprint {http://arxiv.org/abs/arXiv:2312.00719} {arXiv:arXiv:2312.00719 [hep-th]} \BibitemShut {NoStop}%
\bibitem [{\citenamefont {Damour}\ and\ \citenamefont {Nagar}(2009)}]{damour2009}%
  \BibitemOpen
  \bibfield  {author} {\bibinfo {author} {\bibfnamefont {T.}~\bibnamefont {Damour}}\ and\ \bibinfo {author} {\bibfnamefont {A.}~\bibnamefont {Nagar}},\ }\href@noop {} {\bibfield  {journal} {\bibinfo  {journal} {Phys. Rev. D}\ }\textbf {\bibinfo {volume} {80}},\ \bibinfo {pages} {084035} (\bibinfo {year} {2009})},\ \Eprint {http://arxiv.org/abs/arXiv:0906.0096} {arXiv:arXiv:0906.0096 [gr-qc]} \BibitemShut {NoStop}%
\bibitem [{\citenamefont {Binnington}\ and\ \citenamefont {Poisson}(2009)}]{binnington2009}%
  \BibitemOpen
  \bibfield  {author} {\bibinfo {author} {\bibfnamefont {T.}~\bibnamefont {Binnington}}\ and\ \bibinfo {author} {\bibfnamefont {E.}~\bibnamefont {Poisson}},\ }\href@noop {} {\bibfield  {journal} {\bibinfo  {journal} {Phys. Rev. D}\ }\textbf {\bibinfo {volume} {80}},\ \bibinfo {pages} {084018} (\bibinfo {year} {2009})},\ \Eprint {http://arxiv.org/abs/arXiv:0906.1366} {arXiv:arXiv:0906.1366 [gr-qc]} \BibitemShut {NoStop}%
\bibitem [{\citenamefont {Fang}\ and\ \citenamefont {Lovelace}(2005)}]{fang2005}%
  \BibitemOpen
  \bibfield  {author} {\bibinfo {author} {\bibfnamefont {H.}~\bibnamefont {Fang}}\ and\ \bibinfo {author} {\bibfnamefont {G.}~\bibnamefont {Lovelace}},\ }\href@noop {} {\bibfield  {journal} {\bibinfo  {journal} {Phys. Rev. D}\ }\textbf {\bibinfo {volume} {72}},\ \bibinfo {pages} {124016} (\bibinfo {year} {2005})},\ \Eprint {http://arxiv.org/abs/arXiv:gr-qc/0505156} {arXiv:arXiv:gr-qc/0505156 [gr-qc]} \BibitemShut {NoStop}%
\bibitem [{\citenamefont {Kol}\ and\ \citenamefont {Smolkin}(2012)}]{kol2012}%
  \BibitemOpen
  \bibfield  {author} {\bibinfo {author} {\bibfnamefont {B.}~\bibnamefont {Kol}}\ and\ \bibinfo {author} {\bibfnamefont {M.}~\bibnamefont {Smolkin}},\ }\href@noop {} {\bibfield  {journal} {\bibinfo  {journal} {JHEP}\ }\textbf {\bibinfo {volume} {02}},\ \bibinfo {pages} {010} (\bibinfo {year} {2012})},\ \Eprint {http://arxiv.org/abs/arXiv:1110.3764} {arXiv:arXiv:1110.3764 [hep-th]} \BibitemShut {NoStop}%
\bibitem [{\citenamefont {Chakrabarti}\ \emph {et~al.}(2013)\citenamefont {Chakrabarti}, \citenamefont {Delsate},\ and\ \citenamefont {Steinhoff}}]{Chakrabarti:2013lua}%
  \BibitemOpen
  \bibfield  {author} {\bibinfo {author} {\bibfnamefont {S.}~\bibnamefont {Chakrabarti}}, \bibinfo {author} {\bibfnamefont {T.}~\bibnamefont {Delsate}}, \ and\ \bibinfo {author} {\bibfnamefont {J.}~\bibnamefont {Steinhoff}},\ }\href@noop {} {\bibfield  {journal} {\bibinfo  {journal} {.}\ } (\bibinfo {year} {2013})},\ \Eprint {http://arxiv.org/abs/1304.2228} {arXiv:1304.2228 [gr-qc]} \BibitemShut {NoStop}%
\bibitem [{\citenamefont {Gürlebeck}(2015)}]{gurlebeck2015}%
  \BibitemOpen
  \bibfield  {author} {\bibinfo {author} {\bibfnamefont {N.}~\bibnamefont {Gürlebeck}},\ }\href@noop {} {\bibfield  {journal} {\bibinfo  {journal} {Phys. Rev. Lett.}\ }\textbf {\bibinfo {volume} {114}},\ \bibinfo {pages} {151102} (\bibinfo {year} {2015})},\ \Eprint {http://arxiv.org/abs/arXiv:1503.03240} {arXiv:arXiv:1503.03240 [gr-qc]} \BibitemShut {NoStop}%
\bibitem [{\citenamefont {Porto}(2016)}]{porto2016}%
  \BibitemOpen
  \bibfield  {author} {\bibinfo {author} {\bibfnamefont {R.~A.}\ \bibnamefont {Porto}},\ }\href@noop {} {\bibfield  {journal} {\bibinfo  {journal} {Fortsch. Phys.}\ }\textbf {\bibinfo {volume} {64}},\ \bibinfo {pages} {723–729} (\bibinfo {year} {2016})},\ \Eprint {http://arxiv.org/abs/arXiv:1606.08895} {arXiv:arXiv:1606.08895 [gr-qc]} \BibitemShut {NoStop}%
\bibitem [{\citenamefont {Tiec}\ and\ \citenamefont {Casals}(2021)}]{letiec2021a}%
  \BibitemOpen
  \bibfield  {author} {\bibinfo {author} {\bibfnamefont {A.~L.}\ \bibnamefont {Tiec}}\ and\ \bibinfo {author} {\bibfnamefont {M.}~\bibnamefont {Casals}},\ }\href@noop {} {\bibfield  {journal} {\bibinfo  {journal} {Phys. Rev. Lett.}\ }\textbf {\bibinfo {volume} {126}},\ \bibinfo {pages} {131102} (\bibinfo {year} {2021})},\ \Eprint {http://arxiv.org/abs/arXiv:2007.00214} {arXiv:arXiv:2007.00214 [gr-qc]} \BibitemShut {NoStop}%
\bibitem [{\citenamefont {Tiec}\ \emph {et~al.}(2021)\citenamefont {Tiec}, \citenamefont {Casals},\ and\ \citenamefont {Franzin}}]{letiec2021b}%
  \BibitemOpen
  \bibfield  {author} {\bibinfo {author} {\bibfnamefont {A.~L.}\ \bibnamefont {Tiec}}, \bibinfo {author} {\bibfnamefont {M.}~\bibnamefont {Casals}}, \ and\ \bibinfo {author} {\bibfnamefont {E.}~\bibnamefont {Franzin}},\ }\href@noop {} {\bibfield  {journal} {\bibinfo  {journal} {Phys. Rev. D}\ }\textbf {\bibinfo {volume} {103}},\ \bibinfo {pages} {084021} (\bibinfo {year} {2021})},\ \Eprint {http://arxiv.org/abs/arXiv:2010.15795} {arXiv:arXiv:2010.15795 [gr-qc]} \BibitemShut {NoStop}%
\bibitem [{\citenamefont {Chia}(2021{\natexlab{a}})}]{chia2021}%
  \BibitemOpen
  \bibfield  {author} {\bibinfo {author} {\bibfnamefont {H.~S.}\ \bibnamefont {Chia}},\ }\href@noop {} {\bibfield  {journal} {\bibinfo  {journal} {Phys. Rev. D}\ }\textbf {\bibinfo {volume} {104}},\ \bibinfo {pages} {024013} (\bibinfo {year} {2021}{\natexlab{a}})},\ \Eprint {http://arxiv.org/abs/arXiv:2010.07300} {arXiv:arXiv:2010.07300 [gr-qc]} \BibitemShut {NoStop}%
\bibitem [{\citenamefont {Charalambous}\ \emph {et~al.}(2021{\natexlab{a}})\citenamefont {Charalambous}, \citenamefont {Dubovsky},\ and\ \citenamefont {Ivanov}}]{charalambous2021}%
  \BibitemOpen
  \bibfield  {author} {\bibinfo {author} {\bibfnamefont {P.}~\bibnamefont {Charalambous}}, \bibinfo {author} {\bibfnamefont {S.}~\bibnamefont {Dubovsky}}, \ and\ \bibinfo {author} {\bibfnamefont {M.~M.}\ \bibnamefont {Ivanov}},\ }\href@noop {} {\bibfield  {journal} {\bibinfo  {journal} {JHEP}\ }\textbf {\bibinfo {volume} {05}},\ \bibinfo {pages} {038} (\bibinfo {year} {2021}{\natexlab{a}})},\ \Eprint {http://arxiv.org/abs/arXiv:2102.08917} {arXiv:arXiv:2102.08917 [hep-th]} \BibitemShut {NoStop}%
\bibitem [{\citenamefont {Riva}\ \emph {et~al.}(2023)\citenamefont {Riva}, \citenamefont {Santoni}, \citenamefont {Savi\'c},\ and\ \citenamefont {Vernizzi}}]{Riva:2023rcm}%
  \BibitemOpen
  \bibfield  {author} {\bibinfo {author} {\bibfnamefont {M.~M.}\ \bibnamefont {Riva}}, \bibinfo {author} {\bibfnamefont {L.}~\bibnamefont {Santoni}}, \bibinfo {author} {\bibfnamefont {N.}~\bibnamefont {Savi\'c}}, \ and\ \bibinfo {author} {\bibfnamefont {F.}~\bibnamefont {Vernizzi}},\ }\href@noop {} {\bibfield  {journal} {\bibinfo  {journal} {.}\ } (\bibinfo {year} {2023})},\ \Eprint {http://arxiv.org/abs/2312.05065} {arXiv:2312.05065 [gr-qc]} \BibitemShut {NoStop}%
\bibitem [{\citenamefont {Charalambous}\ \emph {et~al.}(2021{\natexlab{b}})\citenamefont {Charalambous}, \citenamefont {Dubovsky},\ and\ \citenamefont {Ivanov}}]{Charalambous:2021mea}%
  \BibitemOpen
  \bibfield  {author} {\bibinfo {author} {\bibfnamefont {P.}~\bibnamefont {Charalambous}}, \bibinfo {author} {\bibfnamefont {S.}~\bibnamefont {Dubovsky}}, \ and\ \bibinfo {author} {\bibfnamefont {M.~M.}\ \bibnamefont {Ivanov}},\ }\href {\doibase 10.1007/JHEP05(2021)038} {\bibfield  {journal} {\bibinfo  {journal} {JHEP}\ }\textbf {\bibinfo {volume} {05}},\ \bibinfo {pages} {038} (\bibinfo {year} {2021}{\natexlab{b}})},\ \Eprint {http://arxiv.org/abs/2102.08917} {arXiv:2102.08917 [hep-th]} \BibitemShut {NoStop}%
\bibitem [{\citenamefont {Hui}\ \emph {et~al.}(2022{\natexlab{a}})\citenamefont {Hui}, \citenamefont {Joyce}, \citenamefont {Penco}, \citenamefont {Santoni},\ and\ \citenamefont {Solomon}}]{Hui:2021vcv}%
  \BibitemOpen
  \bibfield  {author} {\bibinfo {author} {\bibfnamefont {L.}~\bibnamefont {Hui}}, \bibinfo {author} {\bibfnamefont {A.}~\bibnamefont {Joyce}}, \bibinfo {author} {\bibfnamefont {R.}~\bibnamefont {Penco}}, \bibinfo {author} {\bibfnamefont {L.}~\bibnamefont {Santoni}}, \ and\ \bibinfo {author} {\bibfnamefont {A.~R.}\ \bibnamefont {Solomon}},\ }\href {\doibase 10.1088/1475-7516/2022/01/032} {\bibfield  {journal} {\bibinfo  {journal} {JCAP}\ }\textbf {\bibinfo {volume} {01}},\ \bibinfo {pages} {032} (\bibinfo {year} {2022}{\natexlab{a}})},\ \Eprint {http://arxiv.org/abs/2105.01069} {arXiv:2105.01069 [hep-th]} \BibitemShut {NoStop}%
\bibitem [{\citenamefont {Chia}(2021{\natexlab{b}})}]{Chia:2020yla}%
  \BibitemOpen
  \bibfield  {author} {\bibinfo {author} {\bibfnamefont {H.~S.}\ \bibnamefont {Chia}},\ }\href {\doibase 10.1103/PhysRevD.104.024013} {\bibfield  {journal} {\bibinfo  {journal} {Phys. Rev. D}\ }\textbf {\bibinfo {volume} {104}},\ \bibinfo {pages} {024013} (\bibinfo {year} {2021}{\natexlab{b}})},\ \Eprint {http://arxiv.org/abs/2010.07300} {arXiv:2010.07300 [gr-qc]} \BibitemShut {NoStop}%
\bibitem [{\citenamefont {Charalambous}\ \emph {et~al.}(2021{\natexlab{c}})\citenamefont {Charalambous}, \citenamefont {Dubovsky},\ and\ \citenamefont {Ivanov}}]{Charalambous:2021kcz}%
  \BibitemOpen
  \bibfield  {author} {\bibinfo {author} {\bibfnamefont {P.}~\bibnamefont {Charalambous}}, \bibinfo {author} {\bibfnamefont {S.}~\bibnamefont {Dubovsky}}, \ and\ \bibinfo {author} {\bibfnamefont {M.~M.}\ \bibnamefont {Ivanov}},\ }\href {\doibase 10.1103/PhysRevLett.127.101101} {\bibfield  {journal} {\bibinfo  {journal} {Phys. Rev. Lett.}\ }\textbf {\bibinfo {volume} {127}},\ \bibinfo {pages} {101101} (\bibinfo {year} {2021}{\natexlab{c}})},\ \Eprint {http://arxiv.org/abs/2103.01234} {arXiv:2103.01234 [hep-th]} \BibitemShut {NoStop}%
\bibitem [{\citenamefont {Charalambous}\ \emph {et~al.}(2022)\citenamefont {Charalambous}, \citenamefont {Dubovsky},\ and\ \citenamefont {Ivanov}}]{Charalambous:2022rre}%
  \BibitemOpen
  \bibfield  {author} {\bibinfo {author} {\bibfnamefont {P.}~\bibnamefont {Charalambous}}, \bibinfo {author} {\bibfnamefont {S.}~\bibnamefont {Dubovsky}}, \ and\ \bibinfo {author} {\bibfnamefont {M.~M.}\ \bibnamefont {Ivanov}},\ }\href {\doibase 10.1007/JHEP10(2022)175} {\bibfield  {journal} {\bibinfo  {journal} {JHEP}\ }\textbf {\bibinfo {volume} {10}},\ \bibinfo {pages} {175} (\bibinfo {year} {2022})},\ \Eprint {http://arxiv.org/abs/2209.02091} {arXiv:2209.02091 [hep-th]} \BibitemShut {NoStop}%
\bibitem [{\citenamefont {Ben~Achour}\ \emph {et~al.}(2022)\citenamefont {Ben~Achour}, \citenamefont {Livine}, \citenamefont {Mukohyama},\ and\ \citenamefont {Uzan}}]{BenAchour:2022uqo}%
  \BibitemOpen
  \bibfield  {author} {\bibinfo {author} {\bibfnamefont {J.}~\bibnamefont {Ben~Achour}}, \bibinfo {author} {\bibfnamefont {E.~R.}\ \bibnamefont {Livine}}, \bibinfo {author} {\bibfnamefont {S.}~\bibnamefont {Mukohyama}}, \ and\ \bibinfo {author} {\bibfnamefont {J.-P.}\ \bibnamefont {Uzan}},\ }\href {\doibase 10.1007/JHEP07(2022)112} {\bibfield  {journal} {\bibinfo  {journal} {JHEP}\ }\textbf {\bibinfo {volume} {07}},\ \bibinfo {pages} {112} (\bibinfo {year} {2022})},\ \Eprint {http://arxiv.org/abs/2202.12828} {arXiv:2202.12828 [gr-qc]} \BibitemShut {NoStop}%
\bibitem [{\citenamefont {Hui}\ \emph {et~al.}(2022{\natexlab{b}})\citenamefont {Hui}, \citenamefont {Joyce}, \citenamefont {Penco}, \citenamefont {Santoni},\ and\ \citenamefont {Solomon}}]{Hui:2022vbh}%
  \BibitemOpen
  \bibfield  {author} {\bibinfo {author} {\bibfnamefont {L.}~\bibnamefont {Hui}}, \bibinfo {author} {\bibfnamefont {A.}~\bibnamefont {Joyce}}, \bibinfo {author} {\bibfnamefont {R.}~\bibnamefont {Penco}}, \bibinfo {author} {\bibfnamefont {L.}~\bibnamefont {Santoni}}, \ and\ \bibinfo {author} {\bibfnamefont {A.~R.}\ \bibnamefont {Solomon}},\ }\href {\doibase 10.1007/JHEP09(2022)049} {\bibfield  {journal} {\bibinfo  {journal} {JHEP}\ }\textbf {\bibinfo {volume} {09}},\ \bibinfo {pages} {049} (\bibinfo {year} {2022}{\natexlab{b}})},\ \Eprint {http://arxiv.org/abs/2203.08832} {arXiv:2203.08832 [hep-th]} \BibitemShut {NoStop}%
\bibitem [{\citenamefont {Perry}\ and\ \citenamefont {Rodriguez}(2023)}]{Perry:2023wmm}%
  \BibitemOpen
  \bibfield  {author} {\bibinfo {author} {\bibfnamefont {M.}~\bibnamefont {Perry}}\ and\ \bibinfo {author} {\bibfnamefont {M.~J.}\ \bibnamefont {Rodriguez}},\ }\href@noop {} {\bibfield  {journal} {\bibinfo  {journal} {.}\ } (\bibinfo {year} {2023})},\ \Eprint {http://arxiv.org/abs/2310.03660} {arXiv:2310.03660 [gr-qc]} \BibitemShut {NoStop}%
\bibitem [{\citenamefont {Saketh}\ \emph {et~al.}(2023)\citenamefont {Saketh}, \citenamefont {Zhou},\ and\ \citenamefont {Ivanov}}]{Saketh:2023bul}%
  \BibitemOpen
  \bibfield  {author} {\bibinfo {author} {\bibfnamefont {M.~V.~S.}\ \bibnamefont {Saketh}}, \bibinfo {author} {\bibfnamefont {Z.}~\bibnamefont {Zhou}}, \ and\ \bibinfo {author} {\bibfnamefont {M.~M.}\ \bibnamefont {Ivanov}},\ }\href@noop {} {\bibfield  {journal} {\bibinfo  {journal} {.}\ } (\bibinfo {year} {2023})},\ \Eprint {http://arxiv.org/abs/2307.10391} {arXiv:2307.10391 [hep-th]} \BibitemShut {NoStop}%
\bibitem [{\citenamefont {Cvetic}\ and\ \citenamefont {Larsen}(2012)}]{cvetic:2011dn}%
  \BibitemOpen
  \bibfield  {author} {\bibinfo {author} {\bibfnamefont {M.}~\bibnamefont {Cvetic}}\ and\ \bibinfo {author} {\bibfnamefont {F.}~\bibnamefont {Larsen}},\ }\href {\doibase 10.1007/JHEP09(2012)076} {\bibfield  {journal} {\bibinfo  {journal} {JHEP}\ }\textbf {\bibinfo {volume} {09}},\ \bibinfo {pages} {076} (\bibinfo {year} {2012})},\ \Eprint {http://arxiv.org/abs/1112.4846} {arXiv:1112.4846 [hep-th]} \BibitemShut {NoStop}%
\bibitem [{\citenamefont {Castro}\ \emph {et~al.}(2010)\citenamefont {Castro}, \citenamefont {Maloney},\ and\ \citenamefont {Strominger}}]{Castro:2010fd}%
  \BibitemOpen
  \bibfield  {author} {\bibinfo {author} {\bibfnamefont {A.}~\bibnamefont {Castro}}, \bibinfo {author} {\bibfnamefont {A.}~\bibnamefont {Maloney}}, \ and\ \bibinfo {author} {\bibfnamefont {A.}~\bibnamefont {Strominger}},\ }\href {\doibase 10.1103/PhysRevD.82.024008} {\bibfield  {journal} {\bibinfo  {journal} {Phys. Rev. D}\ }\textbf {\bibinfo {volume} {82}},\ \bibinfo {pages} {024008} (\bibinfo {year} {2010})},\ \Eprint {http://arxiv.org/abs/1004.0996} {arXiv:1004.0996 [hep-th]} \BibitemShut {NoStop}%
\bibitem [{\citenamefont {Cvetic}\ and\ \citenamefont {Gibbons}(2012)}]{Cvetic:2012tr}%
  \BibitemOpen
  \bibfield  {author} {\bibinfo {author} {\bibfnamefont {M.}~\bibnamefont {Cvetic}}\ and\ \bibinfo {author} {\bibfnamefont {G.~W.}\ \bibnamefont {Gibbons}},\ }\href {\doibase 10.1007/JHEP07(2012)014} {\bibfield  {journal} {\bibinfo  {journal} {JHEP}\ }\textbf {\bibinfo {volume} {07}},\ \bibinfo {pages} {014} (\bibinfo {year} {2012})},\ \Eprint {http://arxiv.org/abs/1201.0601} {arXiv:1201.0601 [hep-th]} \BibitemShut {NoStop}%
\bibitem [{\citenamefont {Cveti\v{c}}\ \emph {et~al.}(2024)\citenamefont {Cveti\v{c}}, \citenamefont {Rodr\'\i{}guez}, \citenamefont {Rodriguez},\ and\ \citenamefont {Varela}}]{Cvetic:2024dvn}%
  \BibitemOpen
  \bibfield  {author} {\bibinfo {author} {\bibfnamefont {M.}~\bibnamefont {Cveti\v{c}}}, \bibinfo {author} {\bibfnamefont {N.~H.}\ \bibnamefont {Rodr\'\i{}guez}}, \bibinfo {author} {\bibfnamefont {M.~J.}\ \bibnamefont {Rodriguez}}, \ and\ \bibinfo {author} {\bibfnamefont {O.}~\bibnamefont {Varela}},\ }\href@noop {} {\  (\bibinfo {year} {2024})},\ \Eprint {http://arxiv.org/abs/2406.10458} {arXiv:2406.10458 [hep-th]} \BibitemShut {NoStop}%
\bibitem [{\citenamefont {Chong}\ \emph {et~al.}(2005)\citenamefont {Chong}, \citenamefont {Cvetic}, \citenamefont {Lu},\ and\ \citenamefont {Pope}}]{Chong:2004na}%
  \BibitemOpen
  \bibfield  {author} {\bibinfo {author} {\bibfnamefont {Z.~W.}\ \bibnamefont {Chong}}, \bibinfo {author} {\bibfnamefont {M.}~\bibnamefont {Cvetic}}, \bibinfo {author} {\bibfnamefont {H.}~\bibnamefont {Lu}}, \ and\ \bibinfo {author} {\bibfnamefont {C.~N.}\ \bibnamefont {Pope}},\ }\href {\doibase 10.1016/j.nuclphysb.2005.03.034} {\bibfield  {journal} {\bibinfo  {journal} {Nucl. Phys. B}\ }\textbf {\bibinfo {volume} {717}},\ \bibinfo {pages} {246} (\bibinfo {year} {2005})},\ \Eprint {http://arxiv.org/abs/hep-th/0411045} {arXiv:hep-th/0411045} \BibitemShut {NoStop}%
\bibitem [{\citenamefont {Castro}\ and\ \citenamefont {Rodriguez}(2012)}]{Castro:2012av}%
  \BibitemOpen
  \bibfield  {author} {\bibinfo {author} {\bibfnamefont {A.}~\bibnamefont {Castro}}\ and\ \bibinfo {author} {\bibfnamefont {M.~J.}\ \bibnamefont {Rodriguez}},\ }\href {\doibase 10.1103/PhysRevD.86.024008} {\bibfield  {journal} {\bibinfo  {journal} {Phys. Rev. D}\ }\textbf {\bibinfo {volume} {86}},\ \bibinfo {pages} {024008} (\bibinfo {year} {2012})},\ \Eprint {http://arxiv.org/abs/1204.1284} {arXiv:1204.1284 [hep-th]} \BibitemShut {NoStop}%
\bibitem [{\citenamefont {Astefanesei}\ \emph {et~al.}(2010)\citenamefont {Astefanesei}, \citenamefont {Mann}, \citenamefont {Rodriguez},\ and\ \citenamefont {Stelea}}]{Astefanesei:2009wi}%
  \BibitemOpen
  \bibfield  {author} {\bibinfo {author} {\bibfnamefont {D.}~\bibnamefont {Astefanesei}}, \bibinfo {author} {\bibfnamefont {R.~B.}\ \bibnamefont {Mann}}, \bibinfo {author} {\bibfnamefont {M.~J.}\ \bibnamefont {Rodriguez}}, \ and\ \bibinfo {author} {\bibfnamefont {C.}~\bibnamefont {Stelea}},\ }\href {\doibase 10.1088/0264-9381/27/16/165004} {\bibfield  {journal} {\bibinfo  {journal} {Class. Quant. Grav.}\ }\textbf {\bibinfo {volume} {27}},\ \bibinfo {pages} {165004} (\bibinfo {year} {2010})},\ \Eprint {http://arxiv.org/abs/0909.3852} {arXiv:0909.3852 [hep-th]} \BibitemShut {NoStop}%
\bibitem [{\citenamefont {Astefanesei}\ \emph {et~al.}(2009)\citenamefont {Astefanesei}, \citenamefont {Rodriguez},\ and\ \citenamefont {Theisen}}]{Astefanesei:2009mc}%
  \BibitemOpen
  \bibfield  {author} {\bibinfo {author} {\bibfnamefont {D.}~\bibnamefont {Astefanesei}}, \bibinfo {author} {\bibfnamefont {M.~J.}\ \bibnamefont {Rodriguez}}, \ and\ \bibinfo {author} {\bibfnamefont {S.}~\bibnamefont {Theisen}},\ }\href {\doibase 10.1088/1126-6708/2009/12/040} {\bibfield  {journal} {\bibinfo  {journal} {JHEP}\ }\textbf {\bibinfo {volume} {12}},\ \bibinfo {pages} {040} (\bibinfo {year} {2009})},\ \Eprint {http://arxiv.org/abs/0909.0008} {arXiv:0909.0008 [hep-th]} \BibitemShut {NoStop}%
\bibitem [{\citenamefont {Cvetic}\ \emph {et~al.}(2015)\citenamefont {Cvetic}, \citenamefont {Gibbons},\ and\ \citenamefont {Saleem}}]{Cvetic:2014nta}%
  \BibitemOpen
  \bibfield  {author} {\bibinfo {author} {\bibfnamefont {M.}~\bibnamefont {Cvetic}}, \bibinfo {author} {\bibfnamefont {G.~W.}\ \bibnamefont {Gibbons}}, \ and\ \bibinfo {author} {\bibfnamefont {Z.~H.}\ \bibnamefont {Saleem}},\ }\href {\doibase 10.1103/PhysRevLett.114.231301} {\bibfield  {journal} {\bibinfo  {journal} {Phys. Rev. Lett.}\ }\textbf {\bibinfo {volume} {114}},\ \bibinfo {pages} {231301} (\bibinfo {year} {2015})},\ \Eprint {http://arxiv.org/abs/1412.5996} {arXiv:1412.5996 [hep-th]} \BibitemShut {NoStop}%
\bibitem [{\citenamefont {Chow}\ and\ \citenamefont {Comp{\`e}re}(2014)}]{chow2014black}%
  \BibitemOpen
  \bibfield  {author} {\bibinfo {author} {\bibfnamefont {D.~D.}\ \bibnamefont {Chow}}\ and\ \bibinfo {author} {\bibfnamefont {G.}~\bibnamefont {Comp{\`e}re}},\ }\href@noop {} {\bibfield  {journal} {\bibinfo  {journal} {Physical Review D}\ }\textbf {\bibinfo {volume} {90}},\ \bibinfo {pages} {025029} (\bibinfo {year} {2014})}\BibitemShut {NoStop}%
\end{thebibliography}%

\end{document}